\documentclass[aps,preprint,epsfig,rotate]{revtex4}
\usepackage{graphicx}
\usepackage{bm}
\usepackage{epsfig}

\input epsf
\begin{document}
\title{Secondary electrons emitted during nuclear $\beta^{-}$ decay in few-electron atoms}

\author{Alexei M. Frolov}
\email[E--mail address: ]{afrolov@uwo.ca}

\affiliation{Department of Applied Mathematics \\
 University of Western Ontario, London, Ontario N6H 5B7, Canada}

\author{David M. Wardlaw}
 \email[E--mail address: ]{dwardlaw@mun.ca}

\affiliation{Department of Chemistry, Memorial University of Newfoundland, St.John's, 
             Newfoundland and Labrador, A1C 5S7, Canada}

\date{\today}

\begin{abstract}

`Additional' ionization of light atoms and ions during nuclear $\beta^{-}$ decay is investigated. The procedure which can be used to determine the corresponding 
transition probabilities and the velocity/energy spectrum of secondary electrons is developed. Emission of very fast secondary electrons ($\delta-$electrons) 
from $\beta^{-}$-decaying atoms is also briefly discussed.  

\noindent 
PACS number(s): 23.40.-s, 12.20.Ds and 14.60.Cd

\end{abstract}

\maketitle
\newpage

\section{Introduction}

As is well known from numerous experiments, nuclear $\beta^{-}-$decay in few- and many-electron atoms often proceeds with an `additional' atomic ionization. The general 
equation of this process can be written as (see, e.g., \cite{Blat}, \cite{Fro05})
\begin{equation}
  X \rightarrow Y^{2+} + e^{-} + e^{-}(\beta) + \overline{\nu} + \Delta E \label{eq1}
\end{equation}
where the symbols $X$ and $Y$ designate two different chemical elements (isotopes) with almost equal masses. The symbols X and Y in Eq.(\ref{eq1}) are used to designate 
both atoms/ions and the corresponding atomic nuclei. If $Q$ is the electric charge of the parent (or incident) nucleus $X$, then the nuclear charge of the final nucleus 
$Y$ is $Q + 1$. Below, the electric charge of the parent nucleus ($Q$) is designated by the notation $Q_1$, while the electric charge of the final nucleus is denoted by 
the notation $Q_2 (= Q + 1)$. In Eq.(\ref{eq1}) the notation $e^{-}$ stands for the secondary (or slow) electron formed in the unbound spectrum during the decay, while  
the notation $e^{-}(\beta)$ designates the primary (or fast) $\beta^{-}-$electron and $\overline{\nu}$ denotes the electron's anti-neutrino. The total energy $\Delta E$ 
released during the $\beta^{-}-$decay, Eq.(\ref{eq1}), is a given value (for each $\beta^{-}-$decay) which cannot be changed in actual experiments. Formally, the numerical 
value of $\Delta E$ coincides with the maximal (kinetic) energy of the primary $\beta^{-}-$electron emitted in Eq.(\ref{eq1}).  

Our goal in this study is the analysis of the properties of secondary electrons emitted during atomic $\beta^{-}-$decay. In general, the properties of secondary electrons, 
e.g., their velocity spectra, can be used to describe the electron density distribution and electron-electron correlations in the incident atom. Moreover, by using recently 
developed experimental methods one can predict many interesting details of $\beta^{-}$ decay in few-electron atoms and ions. Note that despite a number of experiments 
performed to investigate `additional' ionization of atoms during nuclear $\beta^{-}$-decay our current understanding of some important details of this process is still far 
from complete. In particular, the spectrum of the secondary electron emitted during nuclear $\beta^{-}$-decay in atoms has not been investigated in earlier studies. In this 
communication we derive a closed analytical formula for such a spectrum. Furthermore, it is crucial to explain how the electron-electron correlations in parent atoms can 
affect the secondary-electron spectrum. Another interesting problem discussed in this study is the formation of very fast secondary electrons (so-called $\delta-$electrons) 
during nuclear $\beta^{-}$-decay in few-electron atoms/ions.

Since the first papers published in 1950's (see, e.g., \cite{Mig1940}), it became clear that by analyzing numerically generated spectra of the final state probabilities 
during atomic $\beta^{-}-$decay, Eq.(\ref{eq1}), we can obtain a significant amount of useful information about the parent (or incident) atom/ion, including its atomic 
state, presence of various excitations, etc (see, e.g., \cite{Fro05}, \cite{Our1} -  \cite{PRC2}). Furthermore, if the spectra of the final state probabilities could be 
evaluated to high accuracy (from numerical computations), then based on these spectra we would be able to predict the atom and its isotope in which nuclear $\beta^{-}-$decay 
has occured. A number of important details about electron distributions in such atoms/ions can also be accurately predicted. This conclusion is very important in applications 
to few- and many-electron atoms/ions with very short life-times. This emphasizes the importance of knowledge of the final state probabilities for different atoms, ions, 
molecules and atomic clusters.  

In this study we also determine the distributions (or spectra) of the final state probabilities of $\beta^{-}-$decaying atoms/ions, but our main goal is the analysis of the
cases when this decay proceeds with an `additional' atomic ionization, Eq.(\ref{eq1}). Note that currently all calculations of the final state probabilities for 
$\beta^{-}$decaying atoms, ions and molecules are performed with the use of the sudden approximation which is based on the fact that velocities of $\beta^{-}$-electrons 
($v_{\beta}$) emitted during the nuclear $\beta^{-}-$decay are significantly larger than the usual velocities of atomic electrons $v_a$. In particular, in light atoms we 
have $v_{\beta} \ge 50 v_a - 200 v_a$. This is also true for the velocities of the secondary electrons $e^{-}$ which can be emitted as `free' particles during the reaction, 
Eq.(\ref{eq1}), i.e. $v_{\beta} \gg v_{\delta}$. The inequality $v_{\beta} \gg v_a$ allows one to apply the sudden approximation and analyze the nuclear $\beta^{-}$-decay in 
light atoms by calculating the overlaps of the incident and final (non-relativistic) atomic wave functions. The sudden approximation is based on the assumption that the wave 
function of the incident system does not change during the fast process, i.e. its amplitude and phase do not change. In other words, the electron density distribution in the 
maternal atom does not change during the nuclear $\beta^{-}$-decay (see discussions in \cite{LLQ} and \cite{MigK}). 

Our analysis of the properties of secondary electrons emitted during nuclear $\beta^{-}$-decay in few-electron atoms begins from the general discussion of the final state 
probabilities and sudden approximation which has been extensively used in calculations of such probabilities. This problem is discussed in the next Section. In Section III we 
determine the actual velocity spectrum of the secondary $\beta^{-}-$electrons emitted during nuclear $\beta^{-}$-decay of the one-electron tritium atom. The more general case 
of few-electron atoms is considered in Section IV where we show explicitly that the energy/velocity spectra of secondary electrons essentially depend upon electron-electron 
correlations (or, inter-particle correlations) in the parent few-electron atoms/ions. In Section V we evaluate the overall probabilities to observe very fast secondary 
electrons (or $\delta-$electrons) during nuclear $\beta^{-}$-decay in few-electron atoms. Concluding remarks can be found in the last Section. 
  
\section{Final state probabilities}

In the sudden approximation the final state probability of the process, Eq.(\ref{eq1}), equals the overlap integral of the wave function of the parent atom $X$ and wave 
function of the final double-charged ion $Y^{2+}$ multiplied by the wave function of the outgoing (or `free') electron which has a certain momentum ${\bf p}$. The direction of 
the momentum ${\bf p}$ in space coincides with the direction of motion/propagation of the actual free electron that is observed in experiments. Moreover, at large distances 
each of these free-electron wave functions must be a linear combination of a plane wave and incoming spherical wave. Functions with such asymptotics take the form \cite{Maxim} 
(see also \S 136 in \cite{LLQ})
\begin{eqnarray}
 \phi_{p}(r, {\bf n}_p \cdot {\bf n}_r) = N_f \exp(\frac{\pi}{2} \zeta) \Gamma(1 + \imath \zeta) \; \; {}_1F_1\Bigl(-\imath \zeta, 1, -\imath ({\bf p} \cdot {\bf r} - p r)\Bigr) 
 \exp[\imath ({\bf p} \cdot {\bf r})] \label{Cwave} 
\end{eqnarray}
where $N_f$ is the normalization constant defined below, ${}_1F_1(a, b; z)$ is the confluent hypergeometric function and $\zeta = \frac{Q_2}{a_0 p} = \frac{\alpha Q_2}{\gamma v}$, 
where $a_0$ is the Bohr radius, $\alpha$ is the fine structure constant and $\gamma$ is the Lorentz $\gamma-$factor \cite{Jack} (see below) of the moving electron. The notations 
$p$ and $v$ stand for the momentum and velocity of the outgoing (or `free') electron. Also in this equation the two unit vectors ${\bf n}_p$ and ${\bf n}_r$ are defined as follows 
${\bf n}_p = \frac{{\bf p}}{p}$ and ${\bf n}_r = \frac{{\bf r}}{r}$. There are a number of advantages in using the wave function of the free electron which moves in the Coulomb field 
of the central `bare' nucleus, or positively charged ion in the form of Eq.(\ref{Cwave}). Some of these advantages are discussed in \S 136 of \cite{LLQ}. In particular, the choice
of the $\phi_{p}(r, {\bf n}_p \cdot {\bf n}_r)$ function in the form of Eq.(\ref{Cwave}) directly leads to explicit formulas for the probability amplitudes, i.e. there is no need 
to perform any additional transformations of these values. 

Let us consider nuclear $\beta^{-}$ decay in actual atomic systems. First, consider the $\beta^{-}-$decaying hydrogen (or tritium) atom. The whole process is described by the 
following equation: ${}^{3}$H = ${}^{3}$He$^{2+} + e^{-} + e^{-}(\beta) + \overline{\nu}$. For simplicity, we shall assume that the central atomic nucleus is infinitely heavy. 
Also, in this study we shall assume that all parent (or incident) $\beta^{-}-$decaying atoms were in their ground $1^2s-$states (before $\beta^{-}-$decay). In atomic units, where 
$\hbar = 1, m_e = 1$ and $e = 1$, the ground state wave function of the one-electron, hydrogen-like atom/ion is $\frac{\eta \sqrt{\eta}}{\sqrt{\pi}} \exp(-\eta r)$. In the case of 
$\beta^{-}$-decaying hydrogen/tritium atom we chose $Q_1 = Q = 1$ and $\eta = \frac{Q_1}{a_0}$, while for the final helium ion He$^{2+}$ we have $Q_2 = Q + 1(= 2)$ and $\zeta = 
\frac{Q_2}{a_0 p} = \frac{\alpha Q_2}{\gamma v}$ 

The probability amplitude equals the overlap integral between the $\frac{\eta \sqrt{\eta}}{\sqrt{\pi}} \exp(-\eta r)$ function and the $N_f \phi_{kl}(r, {\bf n}_p \cdot {\bf n}_r)$ 
functions, Eq.(\ref{Cwave}). Calculations of similar integrals (or probability amplitudes) with the 
function $\phi_{kl}(r, {\bf n}_p \cdot {\bf n}_r)$, Eq.(\ref{Cwave}), are relatively simple and straightforward. There are a few steps in this procedure. First, we can write the 
following expression derived in \cite{Maxim}
\begin{eqnarray}
 I_1(\eta) &=& 4 \pi \int \exp[\imath ({\bf p} \cdot {\bf r} - \eta r)] \; \; {}_1F_1\Bigl(-\imath \zeta, 1, -\imath ({\bf p} \cdot {\bf r} - p r)\Bigr) r dr \nonumber \\
 &=& 4 \pi \frac12 \Bigl[ \frac12 p^2 + \frac12 \eta^2 \Bigr]^{\imath \zeta - 1} \Bigl[ - \frac12 p^2 + \frac12 \eta^2 - \imath \eta p \Bigr]^{-\imath \zeta} \label{Max1} 
\end{eqnarray}
after a few steps of additional transformations this formula is reduced to the form
\begin{eqnarray}
 I_1(\eta) = 4 \pi \Bigl( \frac{\eta + \imath p}{\eta - \imath p} \Bigr)^{\imath \zeta} \; \; \frac{1}{\eta^2 + p^2}  \label{Max2}
\end{eqnarray}
By using the following identity (see, e.g., Eq.(1.622) in \cite{GR})
\begin{eqnarray}
   \ln\Bigl( \frac{\eta + \imath p}{\eta - \imath p} \Bigr) = 2 \imath \arctan\Bigl(\frac{\eta}{p}\Bigr) \label{Max3}  
\end{eqnarray}
we reduce the expression for the $I_1(\eta)$ integral to the form
\begin{eqnarray}
   I_1(\eta) = 4 \pi \frac{1}{\eta^2 + p^2} \exp\Bigl[-2 \zeta \arctan\Bigl(\frac{\eta}{p}\Bigr)\Bigr] \label{Max4}
\end{eqnarray}
All integrals which are needed to determine amplitudes of the final state probabilities can be derived by calculating partial derivatives of the $I_1(\eta)$ integral, Eq.(\ref{Max4}), with 
respect to the variable $- \eta$. For instance, for our present purposes we need the integral $I_2(\eta)$ which is written in the form
\begin{eqnarray}
  & & I_2(\eta) = 4 \pi \int \exp[\imath ({\bf p} \cdot {\bf r} - \eta r)] \; \; {}_1F_1\Bigl(-\imath \zeta, 1, -\imath ({\bf p} \cdot {\bf r} - p r)\Bigr) r^{2} dr \nonumber \\
  &=& -\frac{\partial I_1(\eta)}{\partial \eta} = 8 \pi \frac{\eta + \zeta p}{(\eta^2 + p^2)^2} \exp\Bigl[-2 \zeta \arctan\Bigl(\frac{\eta}{p}\Bigr)\Bigr]  \label{Max5}
\end{eqnarray}
The $I_2(\eta)$ integral, Eq.(\ref{Max5}) (with the additional normalization factors $N_f$ and $N_{{\rm H}}$) determines the probability of the `additional' ionization of the 
hydrogen/tritium atom from its ground $1^2s$-state during the nuclear $\beta^{\pm}$ decay. The momentum of the `free' electron is ${\bf p}$ and $p = \mid {\bf p} \mid$ is its 
absolute value. If we want to determine the final state probabilities of atomic ionization during nuclear $\beta^{\pm}$ decay of the hydrogen/tritium atom from excited $s-$states, 
then higher derivatives from the $I_1(\eta)$ integral are needed. In general, all integrals $I_n(\eta)$ can be found with the use of the formula
\begin{eqnarray}
  I_n(\eta) = (-1)^{n} \Bigl[\frac{\partial}{\partial \eta}\Bigr]^n I_1(\eta) = 2^{n+2} \; \; \pi \; \; \frac{P_n(\eta, \zeta, p)}{(\eta^2 + p^2)^n} \exp\Bigl[-2 \zeta 
  \arctan\Bigl(\frac{\eta}{p}\Bigr)\Bigr] \label{Max51}
\end{eqnarray}
where $P_n(\eta, \zeta, p)$ is a polynomial function of all its variables. In derivation of formulas for the integrals $I_n(\eta)$ it is convenient to assume that these three variables 
$\eta, \zeta$ and $p$ are independent of each other. However, to produce actual formulas for the probability amplitudes and final state probabilities we have to take into account 
the following relation between these variables: $\frac{\eta}{p} = \frac{Q_1}{Q_2} \zeta$, or $\zeta p = \frac{Q_2}{Q_1} \eta$. This allows us to write the following expression for 
the integral $I_2(\eta)$
\begin{eqnarray}
  I_2(\eta) = 8 \pi \frac{\eta \Bigl(\frac{Q_2}{Q_1} + 1\Bigr)}{(\eta^2 + p^2)^2} \exp\Bigl[-2 \Bigl(\frac{Q_2 \eta}{Q_1 p}\Bigr) \arctan\Bigl(\frac{Q_2 \eta}{Q_1 p} \Bigr)\Bigr]  
 \label{Max53}
\end{eqnarray}
where we have used the two variables $\eta$ and $p$. However, in some cases two other variables (e.g., $\zeta $ and $p$) are more convenient. Note that it is possible to produce a few 
useful relations between $I_n(\eta)$ and $I_{n-1}(\eta), I_{n-2}(\eta), \ldots, I_{1}(\eta)$ integrals. Such relations allow one to determine all integrals $I_{n}(\eta)$ without any 
actual computation. 

\section{Tritium atom}

Consider nuclear $\beta^{-}$-decay in the one-electron hydrogen/tritium atom ${}^{3}$H, or in some hydrogen-like ion with nuclear electric charge $Q$. According to the formulas derived 
above the probability amplitude ${\cal A}_{i \rightarrow f}$ is
\begin{eqnarray}
  {\cal A}_{i \rightarrow f} = 8 \pi N_H N_f \; \; \frac{\eta \Bigl(\frac{Q_2}{Q_1} + 1\Bigr)}{(\eta^2 + p^2)^2} \; \; \exp\Bigl[-2 \Bigl(\frac{Q_2 \eta}{Q_1 p}\Bigr) 
  \arctan\Bigl(\frac{Q_2 \eta}{Q_1 p}\Bigr)\Bigr] 
  \label{Max54}
\end{eqnarray}
where $N_{{\rm H}} = \sqrt{\frac{\eta^3}{\pi a^{3}_{0}}}$ is the normalization constant of the hydrogen-atom wave function, while $N_f$ is the normalization constant of the wave function which 
represent the `free' electron. The numerical value of this normalization constant ($N_f$) is determined by the following equality 
\begin{eqnarray}
  N^{-2}_f = \exp(\frac{\pi}{2} \zeta) \; \Gamma(1 + \imath \zeta) \; \exp(\frac{\pi}{2} \zeta) \Gamma(1 - \imath \zeta) = \exp(\pi \zeta) \frac{\pi \zeta}{\sinh(\pi \zeta)}
  = \frac{2 \pi \zeta}{1 - \exp(-2 \pi \zeta)} \label{Max55}
\end{eqnarray}
see, e.g., \cite{AS}. In other words, the probability amplitude ${\cal A}_{i \rightarrow f}$ equals
\begin{eqnarray}
 {\cal A} = 4 \sqrt{\frac{2 \eta^{3}}{\zeta} \Bigl[1 - \exp\Bigl(-2 \pi \frac{Q_2 \eta}{Q_1 p}\Bigr)\Bigr]} \; \frac{\eta \Bigl(\frac{Q_2}{Q_1} + 1\Bigr)}{(\eta^2 + p^2)^2} \;
 \exp\Bigl[-2 \Bigl(\frac{Q_2 \eta}{Q_1 p}\Bigr) \arctan\Bigl(\frac{Q_2 \eta}{Q_1 p}\Bigr)\Bigr] \label{Max555}
\end{eqnarray}   
The expression for the infinitely small final state probability $\Delta P_{i \rightarrow f}$ takes the form
\begin{eqnarray}
 & & \Delta P_{i \rightarrow f} = \mid {\cal A} \mid^2 p^2 \Delta p = \frac{32 \eta^3}{\zeta} \; \; \Bigl[1 - \exp\Bigl(-2 \pi \frac{Q_2 \eta}{Q_1 p}\Bigr)\Bigr] \; \; 
 \frac{p^2 \eta^2 \Bigl(\frac{Q_2}{Q_1} + 1\Bigr)^2}{(\eta^2 + p^2)^4} \nonumber \\
 & & \times \exp\Bigl[-4 \Bigl(\frac{Q_2 \eta}{Q_1 p}\Bigr) \arctan\Bigl(\frac{Q_2 \eta}{Q_1 p}\Bigr)\Bigr] \Delta p  \label{Max56}
\end{eqnarray} 
To produce the final expression which is ready for calculations we have to replace here the variables $\eta$ and $\zeta$ by the following expressions $\eta = \frac{Q_1}{a_0}, \frac{\eta}{p} 
= \frac{\alpha Q_1}{\gamma v}$ and $\zeta = \frac{Q_2 \eta}{Q_1 p} = \frac{\alpha Q_2}{\gamma v}$, where $Q_1(= Q)$ is the electric charge of the incident nucleus (or central positively charged 
ion) and $a_0 = \frac{\hbar^2}{m_e e^2}$ is the Bohr radius. In atomic units, where $\hbar = 1, e = 1$ and $m_e = 1$, the Bohr radius equals unity and the ratio $\frac{\eta}{p}$ equals to the 
ratio $\frac{\alpha Q_1}{\gamma v}$ (since $m_e = 1)$, where $\alpha = \frac{\hbar^2}{m_e e^2}$ is the fine structure constant and $v = \mid {\bf v} \mid$ is the absolute value of the electron's 
velocity (expressed in atomic units). The factor $\gamma = \frac{1}{\sqrt{1 - \frac{v^2}{c^2}}} = \frac{1}{\sqrt{1 - \alpha^2 v^2}}$ is the Lorentz $\gamma-$factor \cite{Jack} of the moving 
electron. In atomic units the electron's velocity cannot exceed the value of $c = \alpha^{-1} (\approx 137)$.  

\subsection{Velocity spectrum}

From, Eq.(\ref{Max56}), one finds the following expression for the final state probability disctribution, or $P_{i \rightarrow f}(v)$ distribution:
\begin{eqnarray}
 & & \frac{d P_{i \rightarrow f}}{dv} = \frac{32 Q_1}{\alpha Q_2} \; \; \Bigl[1 - \exp\Bigl(-2 \pi \frac{\alpha Q_2}{\gamma v}\Bigr)\Bigr] \; \; 
 \frac{(Q^{2}_{1} + Q^{2}_{2})^2 \gamma^{4} v^{3}}{(Q^{2}_1 + \gamma^2 v^2)^4} \nonumber \\
 & & \times \exp\Bigl[-4 \Bigl(\frac{\alpha Q_2}{\gamma v}\Bigr) \arctan\Bigl(\frac{\alpha Q_2}{\gamma v}\Bigr)\Bigr] \label{Max565}
\end{eqnarray} 
The expression on the right-hand side of this equality essentially coincides with the $v-$spectrum of the `free' electrons emitted during nuclear $\beta^{-}-$decay in one-electron atoms/ions.    
Rigorously speaking, any spectral function must be normalized, i.e. its integral over $v$ (from $v_{min} = 0$ to $v_{max} = c = \alpha^{-1}$ in $a.u.$) must be equal unity. This allows 
one to obtain the following expression for the $v-$spectral function (or $v-$spectrum, for short) \cite{Fro2015}:
\begin{eqnarray}
 & & S_e(v; Q) = \frac{32 Q_1}{{\cal S}(Q) \alpha Q_2} \; \Bigl[1 - \exp\Bigl(-2 \pi \frac{\alpha Q_2}{\gamma v}\Bigr)\Bigr] \; \; \frac{(Q^{2}_{1} + Q^{2}_{2})^2 \gamma^{4} v^{3}}{(Q^{2}_1 
  + \gamma^2 v^2)^4} \nonumber \\
 & & \times \exp\Bigl[-4 \Bigl(\frac{\alpha Q_2}{\gamma v}\Bigr) \arctan\Bigl(\frac{\alpha Q_2}{\gamma v}\Bigr)\Bigr] \label{Max567}
\end{eqnarray} 
where the normalization constant ${\cal S}(Q)$ can be obtained (for each pair $Q_1 = Q$ and $Q_2 = Q + 1$) with the use of numerical integration. For the tritium atom $Q_1 = 1$ and $Q_2 = 2$ we have 
found that ${\cal S}(Q)$ $\approx$ 196.611833628395. As expected, the formula, Eq.(\ref{Max567}), contains only the absolute values of the free-electron velocity $v$ (or momentum $p$) and electric 
charges of the atomic nuclei $Q_1 = Q$ and $Q_2 = Q + 1$. The velocity of the fast $\beta^{-}-$electron is not included in this formula. This is a direct consequence of the sudden approximation 
used to derive this formula. In general, by using the known $v$-spectral function we can evaluate the probability $p(v)$ to observe a secondary electron which moves with the 
velocity $v$ (expressed in atomic units). 

Note that equation (\ref{Max567}) is written in a manifestly relativistic form, i.e. formally the energies of secondary electrons can be arbitrary. However, both wave functions used in our 
calculations of the overlap integrals are non-relativistic. Furthermore, in applications to actual atoms and ions, the total energies of the emitted secondary electrons are non-relativistic, e.g., 
$E \le 50$ $keV$ for arbitrary atoms and $E \le 25$ $keV$ for light atoms. This means we do not need to apply any relativistic, or even semi-relativistic approximation. In other words, we can 
always assume that $\gamma = 1$ in the formula, Eq.(\ref{Max567}). The non-relativistic spectral function of secondary electrons then takes the form
\begin{eqnarray}
 & & S_e(v; Q) = \frac{32 Q_1}{{\cal S}(Q) \alpha Q_2} \; \; \Bigl[1 - \exp\Bigl(-2 \pi \frac{\alpha Q_2}{v}\Bigr)\Bigr] \; \; \frac{(Q^{2}_{1} + Q^{2}_{2})^2 v^{3}}{(Q^{2}_1 
  + v^2)^4} \nonumber \\
 & & \times \exp\Bigl[-4 \Bigl(\frac{\alpha Q_2}{v}\Bigr) \arctan\Bigl(\frac{\alpha Q_2}{v}\Bigr)\Bigr] \label{Max5675}
\end{eqnarray}   
In applications to real (light) atoms the differences between these two spectral functions, defined by Eq.(\ref{Max567}) and Eq.(\ref{Max5675}), are very small for all light atoms. This follows 
from the explicit form of the right-hand side of these two equations, which contains an exponential cut-off factor for large velocities/energies. In this study all computational results have been 
determined with the use of the spectral function, Eq.(\ref{Max567}).  

\subsection{Calculations}

In actual experiments the integral of the spectral function $S_e(v; Q)$ between the $v_1$ and $v_2$ values ($v_2 > v_1$) gives one the probability $P(v_1;v_2)$ to detect the `free' electron emitted
during the process, Eq.(\ref{eq1}), with the velocity bounded between $v_1$ and $v_2$. This probability is normalized over all possible free electron velocities. However, in actual experiments, in 
addition to such bound-free transitions we always observe a large number of bound-bound transitons. In this case the problem of determining the absolute values of probabilities of the partial 
bound-free transitions is reduced to calculations of the conditional probabilities. To solve this problem one needs to know the total probability of the bound-bound transitions $P_{bb}$ during the 
nuclear $\beta^{-}$-decay. If this value is known, then it is easy to find the total probability of the bound-free transitions $P_{bf}$ = 1 - $P_{bb}$ and absolute value of the partial bound-free 
probability ${\cal P}(v_1;v_2) = P_{bf} P(v_1;v_2) = (1 - P_{bb}) P(v_1;v_2)$     
   
Let us consider the $\beta^{-}$-decay in the one-electron tritium atom ${}^{3}$H (or T). For simplicity, here we restrict our analysis to the $\beta^{-}$-decay of the tritium atom from its ground 
$1^{2}s-$state. Moreover, we shall assume that the atomic nucleus in the hydrogen/tritium atom is infinitely heavy. In general, during the nuclear $\beta^{-}$-decay in such a one-electron tritium 
atom one can observe a large number of bound-bound transitions such as H$(1^{2}s) \rightarrow$ He$^{+}(n^{2}s)$, where $n$ is the principal quantum number of the one-electron (or hydrogen-like) 
He$^{+}$ ion. The sudden approximation leads to the conservation of the electron angular momentum (or $L(L + 1)$ value) during nuclear $\beta^{-}$-decays in few-electron atoms. The total electron 
spin (or $S(S + 1)$ value) is also conserved (as well as the spatial parities $\hat{\pi}$ of the incident and final wave functions) \cite{Fro05}. This means that bound-bound transitions from the 
$1^{2}s$-state of the tritium atom to all bound $n^{2}s-$states of the one-electron helium ion (He$^{+}$) are possible. In this study the probabilities of such transitions have been determined to 
high accuracy and can be found in Table I. Their numerical calculations are relatively simple, since we only need to determine the overlap of the two hydrogen-like, i.e., one-electron, wave 
functions. The sum of such probabilities convergences to the total probability of the bound-bound transition. The convergence of the $P_{bb}$ probability obtained with the use of the 100 - 1500 
lowest $n^{2}s-$states in the He$^{+}$ ion can be understood from Table II. The difference between unity and this probability $P_{bb} \approx$ 0.97372735(10) equals the total probability $P_{bf} 
\approx$ 0.02627265(10) of the bound-free transitions for the process, Eq.(\ref{eq1}).In other words, the $P_{bf}$ value is the total ionization probability of the He$^{+}$ ion during nuclear 
$\beta^{-}$-decay in the tritium atom. For the one-electron ${}^3$H atom such a probability ($\approx$ 2.627 \%) is quite small, but in many atoms the probabilities of similar processes are larger. 
For instance, for the $\beta^{-}$-decay of the Li atom from its ground $2^{2}S-$state, the corresponding probability is $\approx$ 15 \% \cite{Our1}. In many weakly-bound atomic ions, e.g., in the 
two-electron H$^{-}$ ion \cite{Fro05}, the overall probability of bound-free transitons is comparable and even larger than the total probability of bound-bound transitions. Numerical calculations 
of the bound-bound state probabilities for other atomic and molecular systems can be found, e.g., in \cite{PRC1}, \cite{PRC2}. Here we do not want to discuss such calculations, since our current 
goal is to investigate the bound-free transitions during nuclear $\beta^{-}$ decay in few-electron atoms.    

Convergence of the spectral integral ${\cal S}(Q)$ for the $\beta^{-}$-decay of the hydrogen/tritum atom with an infinitely heavy nucleus has been investigated in the following way. First, let us note 
that our method is based on the division of the main velocity interval between $v_{min} = 0$ and $v_{max} = \alpha^{-1}$ into $N$ equal intervals $\delta = \frac{v_{max} - v_{min}}{N}$. To perform 
numerical integration each of these intervals $\delta = \frac{v_{max} - v_{min}}{N}$ is separated into $2^{N_s-2} + 1$ interior sub-intervals which are used in the `extended' trapezoidal method 
\cite{Recp} and \cite{Num}. In our calculations both $N$ and $N_s$ values have been varied, e.g., $N$ = 5000, 1000, $\ldots$ and $N_s$ = 6, 8, 10, 12. Finally, we have determined the resulting numerical 
value of ${\cal S}(Q)$ in Eq.(\ref{Max567}) to high accuracy: ${\cal S}(Q) \approx$ 196.611833628395. This value has been used in all numerical calculations of probabilities.

Table III contains numerical results for probabilities of the bound-free transitions $p_{bf}(0,v)$ during nuclear $\beta^{-}$ decay of the hydrogen/tritium atom with infinitely heavy nucleus. In these 
probabilities the velocities of the final electrons (in $a.u.$) are bounded between $v_1 = 0$ and $v_2 = v$. Note again that these probabilities ($p_{bf}(0,v)$) are the absolute probabilities of the 
bound-free transitions, i.e. all bound-bound transitions are ignored. To obtain the total probabilities of the bound-free transitions the $p_{bf}(0,v)$ values from Table III must be multiplied by the 
factor $P_{bf} \approx$ 0.02627265(10). Then one finds for the overall probability to observe seconday (or `free') electrons following nuclear $\beta^{-}$-decay in atoms with the velocity $v$ bounded 
between $v_1$ and $v_2$ values: $\overline{P}_{bf}(v_1,v_2) = P_{bf} (p_{bf}(0,v_2) - p_{bf}(0,v_1))$. For instance, in the case of nuclear $\beta^{-}$ decay of the hydrogen/tritium atom with infinitely 
heavy nucleus the overall probability to observe the secondary (or `free') electron with the velocity located in the interval $0.6 \le v \le 3.0$ is $\overline{P}_{bf}(v_1,v_2) = P_{bf} \cdot 
(p_{bf}(0,v_2) - p_{bf}(0,v_1)) \approx 0.02627265 \cdot (0.901846525528880670 - 0.0659857766537821459) \approx 0.0219602769$, or 2.196028 \% of all $\beta^{-}$ decays. The first conditional probability 
$p_{bf}(0,v_2)$ corresponds to $v_2 = 3.0$, while the second value $p_{bf}(0,v_1)$ has been determined for $v_1 = 0.6$. Note that for the $\beta^{-}$-decaying tritium atom the velocities of more than 
90 \% of all secondary electrons are located between $v = 0.4$ and $v = 3.2$ (in $a.u.$) This range of velocities of secondary electrons corresponds to the maximum of the $v-$distribution for the ${}^3$H 
$\rightarrow$ ${}^3$He$^{2+}$ + $e^{-} + e^{-}(\beta) + \overline{\nu}$ decay. Probabilities to observe secondary electrons with different velocity distributions can be evaluated analogously by using our 
results from Table III. In many cases it is more convenient to use the (partial) probabilities $p(v_1, v_2)$ defined for proximate numerical values of the two velocities $v_1$ and $v_2$, rather than the 
probabilities $p(0, v_1)$ and $p(0, v_2)$ defined above. The corresponding numerical values of these probabilities $p(v_1, v_2)$ (for $v_1 \ne 0$) can be found in Table IV.   
  
\section{$\beta^{-}$-decays in few-electron atoms}

Our original interest in problems discussed in this study was based on the fact that in actual applications it is often important to know not only the value $P_{bf}$, but also the so-called partial 
probabilities $p_{i \rightarrow {\bf p}}$, where $i$ is the incident state in the parent atom (tritium), while the notation ${\bf p}$ states for the final state of the `free' electron (in momentum 
space) which moves in the field of the final ion (He$^{2+}$ ion). We have developed an effective method for numerical calculations of such probabilities. This method is described in detail below. 
By using the formulas Eq.(\ref{Max565}) and Eq.(\ref{Max567}) one can determine all final state probabilities and $p-$ and $v-$spectra of the secondary (or `free') electrons emitted during nuclear 
$\beta^{-}$-decay in few-electron atoms. In general, our additional investigations of atomic ionization during nuclear $\beta^{-}$-decay in few-electron atoms unambiguously lead to the conclusion that 
the spectra of secondary electrons, partial probabilities of bound-free transitions $p_{i \rightarrow {\bf p}}$, and the total probability of such transitions $P_{bf}$ depend upon the electron-electron 
correlations in the incident bound state of the maternal atom. This means that we can study electron-electron correlations in the maternal (or parent) atom by analyzing the spectra of the secondary 
electrons emitted during its nuclear $\beta^{-}$ decay. This conclusion is important for future experimental studies. 

To illustrate the general situation with few-electron atoms and ions let us consider $\beta-$decaying two-electron atoms and ions, i.e., He-like atomic systems with $\beta^{-}$-decays. Simple and very 
compact analytical expressions for the bound state wave functions of two-electron atoms/ions can be derived in relative and/or perimetric coordinates \cite{Fro98}. The exact wave functions of such 
atomic systems are truly correlated and depend upon all three relative coordinates $r_{32}, r_{31}$ and $r_{21}$. It is very difficult to explain in a few lines all aspects of integration in relative 
and/or perimetric coordinates and we do not attempt to do so here. For our purposes in this study we can operate with the following approximate analytical expression for the two-electron wave function 
(see, e.g, \cite{LLQ} and \cite{March}):
\begin{eqnarray}
  \Psi = N_1 N_2 \; \; \exp[ -(Q - q) (r_{1N} + r_{2N})] = \frac{(Q - q)}{\pi a_0} \; \; \exp[ -(Q - q) (r_{1N} + r_{2N})] \label{two-el}
\end{eqnarray}
where $Q$ is the electric charge of atomic nucleus ($Q \geq 2$), while $Q - q$ is the `effective' electric charge of atomic nucleus. A small correction $q$ ($q \le 1$) is introduced in this equation 
to represent an `effective' contribution of electron-electron correlations. In Eq.(\ref{two-el}) the indexes 1 and 2 stand for the two atomic electrons, while index $N$ designates the atomic nucleus 
whichis assumed to be infinitely heavy. It can be shown that such a simple wave function provides a quite accurate approximation to the actual two-electron wave function. For the ground state of the 
He atom, the approximate wave function, Eq.(\ref{two-el}), reproduces $\approx$ 98.15 \% of its `exact' total energy. The optimal value of the parameter $q$ in Eq.(\ref{two-el}) equals $\frac{5}{16}$ 
\cite{LLQ}, \cite{March}. On the other hand, the approximate wave function is represented in a factorized form (see, Eq.(\ref{two-el})), which contains no mix of inter-electron coordinates. Now, we 
can repeat all calculations made in this study by using the approximate wave function, Eq.(\ref{two-el}). Finally, we arrive at the following expression for the $v-$spectrum of secondary electrons 
emitted during the nuclear $\beta^{-}$ decay of the two-electron atom/ion with the nuclear electric charge $Q$:  
\begin{eqnarray}
 & & S_e(v; Q; q) =  F(Q; q) \frac{32 Q_1}{{\cal S}(Q;q) \alpha Q_2} \; \; \Bigl[1 + \exp\Bigl(-2 \pi \frac{Q_2 \alpha}{\gamma v}\Bigr)\Bigr] \; \; \frac{(Q^{2}_{1} + Q^{2}_{2})^2 \gamma^{4} 
  v^{3}}{(Q^{2}_1 + \gamma^2 v^2)^4} \nonumber \\
 & & \times \exp\Bigl[-4 \Bigl(\frac{\alpha Q_2}{\gamma v}\Bigr) \arctan\Bigl(\frac{\alpha Q_2}{\gamma v}\Bigr)\Bigr] \label{Max568}
\end{eqnarray}
where $Q_1 = Q - q, Q_2 = Q + 1$ and the additional factor $F(Q; q)$ is written in the form
\begin{equation}
  F(Q; q) = \frac{\sqrt{Q^3 (Q - q)^3}}{(Q - \frac{q}{2}\Bigr)^{3}} 
\end{equation}
This factor is, in fact, the probability that the second electron will stay bound (in the ground $1s-$state of the newly formed hydrogen-like ion) during the nuclear $\beta^{-}$-decay in the two-electron 
He-like atom/ion. As one can see from Eq.(\ref{Max568}) the correction for the electron-electron correlations (factor $q$ from Eq.(\ref{two-el})) is included in the final expression for the spectral 
function $S_e(v; Q; q)$, Eq.(\ref{Max568}), of secondary electrons. In addition to the appearence of an extra factor $F(Q;q)$ in Eq.(\ref{Max568}), this factor also changes the `effective' electric charge 
of the nucleus in the incident atom/ion ($Q_1 = Q - q$) and produces changes in the normalization constant ${\cal S}(Q;q)$ in the expression for the spectral function (or spectrum) of secondary electrons. 
These observations illustrate the idea that electron-electron correlations in the maternal atom directly affect the explicit form of the spectra of secondary electrons emitted during the nuclear $\beta^{-}$ 
decay. For few-electron atoms this statement can be rigorously proved with the use of the natural orbital expansions for highly accurate (or truly correlated) variational wave functions for such systems 
(see, e.g., \cite{MCQ}, \cite{David}). Note again that in the non-relativistic approximation we have to assume that $\gamma = 1$ in Eq.(\ref{Max568}) and $v$ is expressed in atomic units, where the unit 
velocity equals the $\frac{e^2}{\hbar} = \alpha c$ value.  

In general, it is hard to determine the final state probabilities in few-electron atoms/ions to the same accuracy as we did above for the one-electron tritium atom. The main problem is related to accurate 
evaluations of the electron-electron correlations in such atomic systems. Another problem in actual calculations of the overlap integrals between the incident and final wave functions follows from the fact 
that the total numbers of essential (or internal) variables are different in these wave functions. For simplicity, let us consider the nuclear $\beta^{-}-$decay of the three electron Li atom which originally 
was in its ground $1^2S-$state. In this case Eq.(\ref{eq1}) takes the form 
\begin{equation}
  {\rm Li} \rightarrow {\rm Be}^{2+} + e^{-} + e^{-}(\beta) + \overline{\nu} \label{eq11}
\end{equation}
Suppose we want to use the bound state wave functions for the incident Li atom and Be$^{2+}$ ion. The incident wave function of the Li-atom contains six inter-particle coordinates, e.g., three 
electron-nucleus coordinates $r_{4i}$ ($i$ = 1, 2, 3) and three electron-electron coordinates $r_{12}, r_{13}, r_{23}$. In the final wave function which describes the Be$^{2+}$ ion and a `free' electron 
one finds three electron-nucleus coordinates $r_{4i}$ ($i$ = 1, 2, 3) and only one electron-electron coordinate $r_{12}$. Here we assume that the `free' electron wave function, Eq.(\ref{Cwave}), depends 
upon the $r_{43} = r_{34}$ electron-nucleus coordinate only. Briefly, this means that the two electron-electron coordinates $r_{13}, r_{23}$ are lost during the sudden transition form the incident to the 
final state in Eq.(\ref{eq11}). In atomic systems with five-, six- and more electrons there are additional problems related to the appearance of the so-called `unnecessary' relative coordinates in the 
bound state wave functions (for more details, see, e.g., \cite{Fro06}). For instance, there are ten relative coordinates (since the number of combinations from 5 by 2 is: $C^{2}_{5}$ = 10) in an arbitrary 
four-electron atom/ion, but only nine of them are truly independent in three-dimensional space. Here we cannot discuss all aspects of these interesting problems and note only that each of these two 
problems presents significant difficulties for accurate computations of actual atoms and ions. 
 
Finally, we have developed an approximate method which can be used to determine the final state probabilities for all states which arise after the nuclear $\beta^{-}$-decay and which belong to the 
continuous spectrum of the final ion, Eq.(\ref{eq11}). This method is based on the natural orbital expansions of all few-electron wave functions which are included in the overlap integral between wave 
functions of the incident and final states. For the process, Eq.(\ref{eq11}), the wave function of the incident state describes the ground $2^2S-$state of the three-electron Li atom. The final state wave 
function is the product of the bound state wave function of the two-electron Be$^{2+}$ ion and the one-electron wave function of the `free' electron, Eq.(\ref{Cwave}) which moves in the central field of 
this ion. In the method of natural orbital expansions the bound state wave functions of few- and many-electron atoms are represented by the sums of the products of their natural orbitals $\chi_{k}(r_{i}) 
= \chi_{k}(r_{iN})$ (the symbol $N$ stands here for the nucleus) which are some simple single-electron functions of one radial variable $r_{iN} = r_{i}$ only. In other words, we are looking for the best 
approximation of the actual wave function of an $N_e-$electron atomic system by linear combinations of $N_e$-products of functions each of which depends upon one radial electron-nucleus coordinate $
r_{iN}$ ($i = 1, \ldots, N_e$) only. The natural orbital expansion is the `best' of all such linear combinations in Dirac's sense \cite{Dirac}, since the first-order density matrix is diagonal in the 
natural orbitals.  

In our case for the three-electron Li-atom and final two-electron Be$^{2+}$ ion we can write the following natural orbital expansions
\begin{eqnarray}
  \Psi_{L=0}(\bigl\{ r_{ij} \bigr\})({\rm Li}) &=& \sum^{N_1}_{n=1} C_n \chi^{(1)}_{n}(r_{1}) \chi^{(2)}_{n}(r_{2}) \chi^{(3)}_{n}(r_{3}) \label{no1} \\
  \Psi_{L=0}(\bigl\{ r_{ij} \bigr\})({\rm Be}^{2+}) &=& \sum^{N_2}_{k=1} B_k \xi^{(1)}_{k}(r_{1}) \xi^{(2)}_{k}(r_{2})  \label{no2} 
\end{eqnarray} 
respectively. Here $\chi_{n}(r_{i})$ and $\xi^{(i)}_{n}(r_{i})$ are the (atomic) natural orbitals constructed for the three-electron Li atom and two-electron Be$^{2+}$ ion (see, e.g., \cite{MCQ}, \cite{David}). 
The coefficients $C_n$ and $B_k$ are the coefficients of the natural orbital expansions constructed for the $2^2S$-state of the Li atom and for the ground $1^1S-$state of the Be$^+$ ion, respectively. In 
general, these coefficients are determined as the solutions (eigenvectors) of associated eigenvalue problems. Note that each of these natural orbitals depends upon the corresponding electron-nucleus coordinate 
$r_{i}$ only (or $r_{4i}$ coordinate in our notation). In general, the natural orbital expansions do not include any of the electron-electron (or correlation) coordinates. The use of the natural orbital 
expansions for the few-electron wave functions allows one to simplify drastically all calculations of the final state probabilities. Indeed, by using the natural orbital expansions one can show that all overlap 
integrals are represented as the product of three one-dimensional integrals, or as finite sums of such products. Briefly, we can say that application of the natural orbital expansions for few-electron atomic 
wave functions allows one to reduce calculations of the overlap integrals to a very simple procedure, e.g., for the process, Eq.(\ref{eq11}), one finds for the probability amplitude $M_{if}$:
\begin{eqnarray}
 M_{if} &=& \sum^{N_1}_{n=1} \sum^{N_2}_{k=1} C_n B_k \int_{0}^{+\infty} \chi^{(1)}_{n}(r_{1}) \xi^{(1)}_{k}(r_{1}) r^2_1 dr_1 \int_{0}^{+\infty} \chi^{(2)}_{n}(r_{2}) \xi^{(2)}_{k}(r_{2}) 
 r^2_2 dr_2 \nonumber \\ 
 & & \times \int_{0}^{+\infty} \chi^{(3)}_{n}(r_{3}) \phi_{kl}(r_3) r^2_3 dr_3 \label{amplt} 
\end{eqnarray}
where $\phi_{kl}(r_3)$ are the functions from Eq.(\ref{Cwave}). In other words, computations of the overlap integrals are now reduced to the calculation of one-dimensional integrals and products of such 
integrals. The total number of integrals used in Eq.(\ref{amplt}) equals the number of bound electrons in the parent (or incident) atom/ion. In other words, in this method we do not face any problem 
related either to different numbers of independent variables in the incident and final wave functions, or to the existance of `unnecessary' relative coordinates in many-electron atomic systems. The formula, 
Eq.(\ref{amplt}), can be used to determine the overall probabilities of the $\beta^{-}$-decay with the emission of a `free' electron during nuclear $\beta^{-}$ decay in three-electron atoms/ions. Analogous 
expressions for the probability amplitudes $M_{if}$ and final state probabilities $P_{if} = \mid M_{if} \mid^2$ can be derived for arbitrary few- and many-electron atoms and ions. 

\section{Formation of fast secondary electrons}

In this Section we briefly discuss the emission of very fast secondary electrons from $\beta^{-}$-decaying few-electron atoms and ions. The velocities of such `fast' secondary electrons significanly exceed 
`averaged' velocities of any `secondary' electron emitted in the process, Eq.(\ref{eq1}). In a number of books and textbooks such fast electrons are often called the $\delta-$electrons. Sudden acceleration 
of these electrons to large velocities is related to the transfering of a large amount of momentum from a very fast, `relativistic' $\beta^{-}$-electron to one of the atomic electrons. Formally, this 
process can be written in the form
\begin{equation}
  X \rightarrow Y^{2+} + e^{-}(\delta) + e^{-}(\beta) + \overline{\nu} \label{fse}
\end{equation}
where $e^{-}(\delta)$ is the fast scondary electron emitted and accelerated to relatively large velocities during nuclear $\beta^{-}$-decay. It is clear that the probability of such a process is small. In the 
lowest-order approximation such a probability is evaluated as $P \approx \alpha^4 P_e$, where $P_e$ is the probability of free-electron emission in the process, Eq.(\ref{fse}), and $\alpha = \frac{e^2}{\hbar 
c} \approx \frac{1}{137}$ is the dimensionless fine-structure constant which is a small numerical value in QED. More accurate evaluation leads to a formula which contains additional factors which increase the 
numerical value of $P$. Let us derive the formula which can be used to evaluate the probability of emission of the fast $\delta-$electrons during $\beta^{-}$-decay in few-electron atoms and ions. 

In reality, the fast secondary electron arises when a substantial amount of momentum-energy is transfered from the very fast $\beta^{-}$-electron to a slow atomic electron. Therefore, we can write the 
following integral relation between the spectral functions of the primary and secondary electrons \cite{Fro2016}
\begin{equation}
   S_{\delta}(\gamma_2) = \int_{1}^{\gamma_{max}} F(\gamma_2, \gamma_1) S_{\beta}(\gamma_1) d\gamma_1 \label{fse1}
\end{equation}
where $S_{\beta}(\gamma_1)$ and $S_{\delta}(\gamma_2)$ are the spectral functions of the primary electrons (or $\beta^{-}$-electrons) and secondary electrons (or $\delta$-electrons), respectively. In this 
equation the notation $F(\gamma_2, \gamma_1)$ stands for the kernel of an integral transformation, which is a real function, if both arguments are bounded between unity and $\alpha^{-1}$. The explicit form 
of this kernel has been found in \cite{Fro2015}. To express this kernel let us introduce the value $\Delta = \frac{\gamma_2  - 1}{\gamma_1  - 1}$, where $\gamma_1$ and $\gamma_2$ are the $\gamma-$factors of 
the $\beta^{-}-$ and $\delta-$electrons, respectively. By using this new variable ($\Delta$) we can write the following formula \cite{Fro2015} for the probability to emit one $\delta-$electron whose 
$\gamma-$factor equals the $\gamma_{2}$ value
\begin{eqnarray}
  P(\gamma_2) = \int_{1}^{\gamma_{max}} \Bigl(\frac{d\sigma}{d \Delta}\Bigr) \Bigl(\frac{d\Delta}{d\gamma_1}\Bigr) S_{\beta}(\gamma_1) d\gamma_1 = (\gamma_2 - 1) \int_{1}^{\gamma_{max}} 
  \Bigl(\frac{d\sigma}{d \Delta}\Bigr) S_{\beta}(\gamma_1) \frac{ d\gamma_1}{(\gamma_1 - 1)^2} \label{fse15}
\end{eqnarray} 
where $\frac{d\Delta}{d\gamma_1} = \frac{\gamma_2 - 1}{(\gamma_1 - 1)^2}$ and the formula for the differential cross-section $\frac{d\sigma}{d\Delta}$ is \cite{Fro2015}:
\begin{eqnarray}
  \frac{d\sigma}{d\Delta} &=& \zeta \frac{16 N_e \pi \alpha^4 a^{2}_{0} \gamma^{2}_1}{(\gamma^{2}_1 - 1) (\gamma_1 - 1)} \; \; \langle \frac{a^{2}_0}{r^{2}_{eN}} \rangle \; \; \frac{1}{\Delta^2 (1 
  - \Delta)^2} \Bigl\{ 1 - \Bigl[ 3 - \Bigl(\frac{\gamma_1 - 1}{\gamma_1}\Bigr)^2 \Bigr] \Delta ( 1 - \Delta) \nonumber \\
  &+& \Bigl(\frac{\gamma_1 - 1}{\gamma_1}\Bigr)^2 \Delta^2 (1 - \Delta)^2 \Bigr\} \label{fse155}
\end{eqnarray}
where $N_e$ is the total number of bound electrons in the parent $\beta^{-}$-decaying atom/ion, $\langle \frac{a^{2}_0}{r^{2}_{eN}} \rangle = \langle \frac{1}{r^{2}_{eN}} \rangle$ (in $a.u.$) is the 
atomic expectation value of $\frac{a^{2}_0}{r^{2}_{eN}} = \frac{1}{r^{2}_{eN}}$ computed for all bound (atomic) electrons, $\zeta$ is some numerical constant, while $\alpha$ and $a_0$ are the 
fine-structure constant and Bohr radius, respectively. Note that the formula, Eq.(\ref{fse155}), can be considered as an integral transformation of the $\beta-$electron spectrum (or spectrum of the 
primary fast electrons). The explicit formula for the spectrum of secondary $\delta-$electrons directly follows from Eqs.(\ref{fse15}) - (\ref{fse155}) which must be integrated over $\gamma_1$ from 1 to 
$\gamma_{max} = \frac{\Delta E}{m_e c^2}$, where $\Delta E$ is the total energy released in the nuclear $\beta^{-}$-decay. This problem can be solved by integrating term-by-term in Eq.(\ref{fse15}), where 
$\frac{d\sigma}{d \Delta}$ must be taken from Eq.(\ref{fse155}).

The final step of our procedure is to find an accurate expression for the spectrum of the primary $\beta^{-}$ electrons which must be used in Eq.(\ref{fse1}). This problem was considered in a large number 
of papers \cite{Fermi} - \cite{Bethe}. Experimental energy spectra of the emited primary $\beta^{-}$ electrons can be found, e.g., in \cite{Cook} and \cite{Neary}, where the $\beta^{-}$ decays of the 
${}^{64}$Cu and ${}^{210}$Bi atoms were studied in detail. As follows from these studies the spectral function of the primary $\beta^{-}$-electrons can be written in the form:
\begin{eqnarray}
  S_{\beta}(\gamma) d\gamma &=& {\cal C}_{\gamma} \cdot F(Q + 1, (\gamma - 1) m_e c^2) \; \; \Bigl[ \frac{\Delta E^{\prime} +  m_e c^{2}}{m_e c^{2}} - \gamma - 1 \Bigr]^2 (\gamma^2 - 1)^{\frac12} 
  \; \; \gamma d\gamma \label{eq55a} \\
   &=& {\cal C}^{\prime}_{\gamma} \cdot F(Q + 1, \gamma - 1) \; \; \Bigl[ \frac{\Delta E^{\prime}}{m_e c^{2}} - \gamma \Bigr]^2 (\gamma^2 - 1)^{\frac12} \; \; \gamma 
  d\gamma \nonumber 
\end{eqnarray} 
where $\Delta E^{\prime} = \Delta E - m_e c^2$. This expression almost exactly coincides with the formula, Eq.(210), derived in \cite{Bethe}, i.e. 
\begin{eqnarray}
  S_{\beta}(\gamma) d\gamma = {\cal C}^{\prime}_{\gamma} \; \; \Bigl[ \frac{\Delta E^{\prime}}{m_e c^{2}} - \gamma \Bigr]^2 (\gamma^2 - 1)^{\frac12} \; \; \gamma d\gamma \label{eq555a} 
\end{eqnarray} 
The spectrum, Eq.(\ref{eq555a}), contains no Fermi function as was introduced by Fermi in \cite{Fermi}. In general, the assumption that $F(Q + 1, \gamma - 1) = 1$ works well for light atoms, but 
for intermediate ($Q \ge 40$) and heavy ($Q \ge 75$) atoms the Fermi function in Eq.(\ref{eq55a}) is really needed. As follows from Eq.(\ref{eq555a}) the normalization constant ${\cal C}^{\prime}_{\gamma}$ 
is a function of the thermal energy released during the nuclear $\beta^{-}$ decay, i.e. of the $\frac{\Delta E^{\prime}}{m_e c^{2}}$ ratio, where $m_e$ = 0.5110998910 $MeV/c^2$. Inverse values of the
normaliztion factors $\Bigl({\cal C}^{\prime}_{\gamma}\Bigr)^{-1}$ determined numerically for different $\Delta E^{\prime}$ values can be found in Table V. By using the formulas, Eqs.(\ref{fse15}) - 
(\ref{fse155}) and Eq.(\ref{eq55a}), one can obtain a closed analytical formula for the probabilities of emission and energy/velocity spectrum of the fast secondary electrons (or $\delta-$electrons) 
emitted during the nuclear $\beta^{-}$ decay in arbitrary few- and many-electron atoms/ions.          

\section{Conclusions}
  
We have considered nuclear $\beta^{-}$-decays in few-electron atoms and ions which lead to an additional ionization of the final ion in which one of the atomic electrons becomes unbound. The procedure is
developed for determining the corresponding  transition probabilities and the velocity/energy spectrum of secondary electrons. Formation of fast secondary electrons ($\delta-$electrons) during 
nuclear $\beta^{-}$-decay in few-electron atoms/ions is also briefly discussed. 

It should be mentioned that the important role of bound-free transitions during the nuclear $\beta^{-}$ decay in few-electron atoms has been emphasized since earlier works by Migdal (see, e.g., \cite{LLQ}, 
\cite{MigK} and references therein). In this study we have chosen the proper wave functions to describe the unbound (or `free') electron which is emitted during the nuclear $\beta^{-}$ decay. This allows 
us to solve a number of long-standing problems, e.g., to derive the explicit formulas for the velocity/energy spectra of secondary electrons emitted during nuclear $\beta^{-}$-decay. Furthermore, now it is 
absolutely clear that the spectra of the emitted secondary electrons have different forms for different few-electron atoms/ions, since these spectra strongly depend upon the electron-electron correlations in 
the bound state of the parent atom/ion. From here one finds the `similarity law' between the velocity spectra of secondary electrons emitted during nuclear $\beta^{-}$-decay of two different atoms/ions 
which have the same (or similar) electron configurations. We also describe an approach which can be useful for derivation of the velocity/energy spectrum of very fast secondary electrons ($\delta-$electrons) 
which are observed during nuclear $\beta^{-}$ decays in few- and many-electron atoms/ions.

\begin{table}[tbp]
   \caption{Probabilities (in \%) of the ground-bound H$(1s) \rightarrow$ He$^{+}(ns)$ transitions during nuclear $\beta^{-}$ decay in the 
            ground $1s$-state of the hydrogen/tritium atom with an infinitely heavy nucleus. $n$ is the principal quantum number of the one-electron 
            He$^{+}$ ion. The notation $s$ corresponds to the states in which electron angular momentum equals zero, i.e. $\ell = 0$.}
     \begin{center}
     \scalebox{0.85}{%
     \begin{tabular}{| c | c | c | c | c | c | c | c |}
      \hline\hline
   $n$  &  $p_{bb}$ & $n$  &  $p_{bb}$ & $n$  &  $p_{bb}$ & $n$  &  $p_{bb}$ \\ 
         \hline
   1 & 0.70233196159122$\times 10^{2}$  & 26 & 0.98497803948971$\times 10^{-3}$ & 51 & 0.12974615801458$\times 10^{-3}$ &  76 & 0.39162862050606$\times 10^{-4}$ \\
   2 & 0.25000000000000$\times 10^{2}$  & 27 & 0.87903331958611$\times 10^{-3}$ & 52 & 0.12239426027150$\times 10^{-3}$ &  77 & 0.37655868762239$\times 10^{-4}$ \\
   3 & 0.12740198400000$\times 10^{1}$  & 28 & 0.78776907687202$\times 10^{-3}$ & 53 & 0.11558764797038$\times 10^{-3}$ &  78 & 0.36225224231066$\times 10^{-4}$ \\
   4 & 0.38536733146295                 & 29 & 0.70872572575441$\times 10^{-3}$ & 54 & 0.10927667532924$\times 10^{-3}$ &  79 & 0.34866153324904$\times 10^{-4}$ \\
   5 & 0.17197881444822                 & 30 & 0.63992249530316$\times 10^{-3}$ & 55 & 0.10341702481201$\times 10^{-3}$ &  80 & 0.33574234908023$\times 10^{-4}$ \\
   6 & 0.92697143554687$\times 10^{-1}$ & 31 & 0.57975202780078$\times 10^{-3}$ & 56 & 0.97969051544095$\times 10^{-4}$ &  81 & 0.32345371594829$\times 10^{-4}$ \\
   7 & 0.55988010386724$\times 10^{-1}$ & 32 & 0.52690070162347$\times 10^{-3}$ & 57 & 0.92897218340823$\times 10^{-4}$ &  82 & 0.31175762421737$\times 10^{-4}$ \\
   8 & 0.36520347436057$\times 10^{-1}$ & 33 & 0.48028782045420$\times 10^{-3}$ & 58 & 0.88169607495337$\times 10^{-4}$ &  83 & 0.30061878124243$\times 10^{-4}$ \\
   9 & 0.25189293972900$\times 10^{-1}$ & 34 & 0.43901878492219$\times 10^{-3}$ & 59 & 0.83757497787194$\times 10^{-4}$ &  84 & 0.29000438743047$\times 10^{-4}$ \\
  10 & 0.18128415546737$\times 10^{-1}$ & 35 & 0.40234872896302$\times 10^{-3}$ & 60 & 0.79634997068789$\times 10^{-4}$ &  85 & 0.27988393315259$\times 10^{-4}$ \\
         \hline
  11 & 0.13491846662324$\times 10^{-1}$ & 36 & 0.36965406024421$\times 10^{-3}$ & 61 & 0.75778722345554$\times 10^{-4}$ &  86 & 0.27022901434745$\times 10^{-4}$ \\
  12 & 0.10317897096839$\times 10^{-1}$ & 37 & 0.34041002208870$\times 10^{-3}$ & 62 & 0.72167520551830$\times 10^{-4}$ &  87 & 0.26101316490293$\times 10^{-4}$ \\
  13 & 0.80702692482061$\times 10^{-2}$ & 38 & 0.31417287990638$\times 10^{-3}$ & 63 & 0.68782224288764$\times 10^{-4}$ &  88 & 0.25221170411742$\times 10^{-4}$ \\
  14 & 0.64331044013362$\times 10^{-2}$ & 39 & 0.29056568635742$\times 10^{-3}$ & 64 & 0.65605437674441$\times 10^{-4}$ &  89 & 0.24380159773154$\times 10^{-4}$ \\
  15 & 0.52118281754238$\times 10^{-2}$ & 40 & 0.26926683590156$\times 10^{-3}$ & 65 & 0.62621348192771$\times 10^{-4}$ &  90 & 0.23576133118693$\times 10^{-4}$ \\
  16 & 0.42819805426639$\times 10^{-2}$ & 41 & 0.25000080828061$\times 10^{-3}$ & 66 & 0.59815561042724$\times 10^{-4}$ &  91 & 0.22807079391474$\times 10^{-4}$ \\
  17 & 0.35613569395414$\times 10^{-2}$ & 42 & 0.23253064079662$\times 10^{-3}$ & 67 & 0.57174953004965$\times 10^{-4}$ &  92 & 0.22071117358569$\times 10^{-4}$ \\
  18 & 0.29941400117151$\times 10^{-2}$ & 43 & 0.21665177430539$\times 10^{-3}$ & 68 & 0.54687543276013$\times 10^{-4}$ &  93 & 0.21366485936703$\times 10^{-4}$ \\
  19 & 0.25415046800868$\times 10^{-2}$ & 44 & 0.20218699709848$\times 10^{-3}$ & 69 & 0.52342379084963$\times 10^{-4}$ &  94 & 0.20691535333240$\times 10^{-4}$ \\
  20 & 0.21758661045733$\times 10^{-2}$ & 45 & 0.18898227105998$\times 10^{-3}$ & 70 & 0.50129434216051$\times 10^{-4}$ &  95 & 0.20044718925958$\times 10^{-4}$ \\
         \hline
  21 & 0.18772530302835$\times 10^{-2}$ & 46 & 0.17690327054347$\times 10^{-3}$ & 71 & 0.48039518821441$\times 10^{-4}$ &  96 & 0.19424585813027$\times 10^{-4}$ \\
  22 & 0.16309610499354$\times 10^{-2}$ & 47 & 0.16583249987351$\times 10^{-3}$ & 72 & 0.46064199130287$\times 10^{-4}$ &  97 & 0.18829773971567$\times 10^{-4}$ \\
  23 & 0.14259985577972$\times 10^{-2}$ & 48 & 0.15566688284309$\times 10^{-3}$ & 73 & 0.44195725848781$\times 10^{-4}$ &  98 & 0.18259003969442$\times 10^{-4}$ \\
  24 & 0.12540360783342$\times 10^{-2}$ & 49 & 0.14631573897880$\times 10^{-3}$ & 74 & 0.42426970206833$\times 10^{-4}$ &  99 & 0.17711073180448$\times 10^{-4}$ \\
  25 & 0.11086825104638$\times 10^{-2}$ & 50 & 0.13769907811358$\times 10^{-3}$ & 75 & 0.40751366744682$\times 10^{-4}$ & 100 & 0.17184850458006$\times 10^{-4}$ \\
         \hline \hline
  \end{tabular}}
  \end{center}
  \end{table}
\begin{table}[tbp]
   \caption{Convergence of the total probabilities $P_{bb}$ of the bound-bound transitions during nuclear $\beta^{-}$ decay of the 
            hydrogen/tritium atom with an infinitely heavy nucleus. $N$ is the total number of hydrogen $ns$-states used in calculations. 
            $n$ is the principal quantum number, while the notation $s$ corresponds to the states in which electron angular momentum 
            $\ell$ equals zero.}
     \begin{center}
     \begin{tabular}{| c | c | c | c | c | c |}
      \hline\hline
 $N$  &  $P_{bb}$ & $N$  &  $P_{bb}$ & $N$  &  $P_{bb}$ \\ 
         \hline
   100 & 0.97371867838323 &  600 & 0.97372694486699 & 1100 & 0.97372711211312 \\
   200 & 0.97372504662813 &  700 & 0.97372700800983 & 1200 & 0.97372712343446 \\
   300 & 0.97372623196518 &  800 & 0.97372704900471 & 1300 & 0.97372713224605 \\
   400 & 0.97372664761376 &  900 & 0.97372707711733 & 1400 & 0.97372713923839 \\
   500 & 0.97372684019166 & 1000 & 0.97372709722987 & 1500 & 0.97372714487989 \\
         \hline \hline
  \end{tabular}
  \end{center}
  \end{table}
\begin{table}[tbp]
   \caption{Probabilities of the bound-free transitions $p_{bf}(0,v)$ during nuclear $\beta^{-}$ decay of the hydrogen/tritium atom with an infinitely 
            heavy nucleus. The velocities of the final electrons (in $a.u.$) are bounded between $v_1 = 0$ and $v_2 = v$.}
     \begin{center}
     \begin{tabular}{| c | c | c | c | c | c |}
      \hline\hline
   $v$  &   $p_{bf}(0,v)$ & $v$  &   $p_{bf}(0,v)$ & $v$  &   $p_{bf}(0,v)$ \\ 
         \hline
     0.2 & 0.202955922212782602$\times 10^{-2}$ &  5.2 & 0.987931066690482215 &  11.0 & 0.999621008255497052 \\
     0.4 & 0.200048139756265682$\times 10^{-1}$ &  5.4 & 0.989736091238550430 &  12.0 & 0.999752250217591601 \\
     0.6 & 0.659857766537821459$\times 10^{-1}$ &  5.6 & 0.991234205230340430 &  13.0 & 0.999832921059878887 \\
     0.8 & 0.141004332183129327 &  5.8 & 0.992483168360981649 &  14.0 & 0.999884258264978038 \\
     1.0 & 0.237287332714349043 &  6.0 & 0.993528952322103761 &  15.0 & 0.999917927291765081 \\
     1.2 & 0.343391270067291913 &  6.2 & 0.994408306447324685 &  16.0 & 0.999940598633865546 \\
     1.4 & 0.448732996357998742 &  6.4 & 0.995150736609713835 &  17.0 & 0.999956223482800655 \\
     1.6 & 0.545899125149580895 &  6.6 & 0.995780036319943809 &  18.0 & 0.999967216325934628 \\
     1.8 & 0.630960454661736217 &  6.8 & 0.996315475460822611 &  19.0 & 0.999975094027025284 \\
     2.0 & 0.702727742898327263 &  7.0 & 0.996772726630822888 &  20.0 & 0.999980833420389250 \\
     2.2 & 0.761747499185994498 &  7.2 & 0.997164589802220096 &  21.0 & 0.999985077730389047 \\
     2.4 & 0.809452784832214921 &  7.4 & 0.997501561449582277 &  22.0 & 0.999988259111051570 \\
     2.6 & 0.847587005336058183 &  7.6 & 0.997792283321484132 &  23.0 & 0.999990673256248918 \\
     2.8 & 0.877871771087206655 &  7.8 & 0.998043897732244535 &  24.0 & 0.999992525883633287 \\
     3.0 & 0.901846525528880670 &  8.0 & 0.998262329974406395 &  25.0 & 0.999993962308590595 \\
     3.2 & 0.920812454828480454 &  8.2 & 0.998452513694625161 &  26.0 & 0.999995086624120623 \\
     3.4 & 0.935832170786307351 &  8.4 & 0.998618571459107597 &  27.0 & 0.999995974360824529 \\
     3.6 & 0.947754919518541073 &  8.6 & 0.998763959977849436 &  28.0 & 0.999996680980098227 \\
     3.8 & 0.957250346302898296 &  8.8 & 0.998891587348680657 &  29.0 & 0.999997247658096693 \\
     4.0 & 0.964842246540587289 &  9.0 & 0.999003908064542578 &  30.0 & 0.999997705279853880 \\
     4.2 & 0.970938568648528298 &  9.2 & 0.999103000282010388 &  35.0 & 0.999999010082609818 \\
     4.4 & 0.975856499673367406 &  9.4 & 0.999190628886819629 &  40.0 & 0.999999532973409329 \\
     4.6 & 0.979842702466459428 &  9.6 & 0.999268297146016705 &  50.0 & 0.999999875793757797 \\
     4.8 & 0.983089287546310995 &  9.8 & 0.999337289155711966 &  75.0 & 0.999999992734260541 \\
     5.0 & 0.985746248754140539 & 10.0 & 0.999398704839936441 & 100.0 & 0.999999999611764017 \\
        \hline \hline
  \end{tabular}
  \end{center}
  \end{table}
\begin{table}[tbp]
   \caption{Probabilities of the bound-free transitions $p_{bf}(v_1,v_2)$ during the nuclear $\beta^{-}$ decay of the 
            tritium atom with an infinitely heavy nucleus. Calculations are performed with the use of the formula, 
            Eq.(\ref{Max567}), where $0 \le v \le \alpha^{-1}$. To obtain the absolute final state probabilities these 
            values must be multiplied by the additional factor $P_{bf} \approx 0.02627265(10)$.}
     \begin{center}
     \scalebox{0.80}{%
     \begin{tabular}{| c | c | c | c | c | c | c | c | c |}
      \hline\hline
 $v_1$ & $v_2$ & $p_{bf}(v_1,v_2)$ &  $v_1$ & $v_2$ & $p_{bf}(v_1,v_2)$ &  $v_1$ & $v_2$ & $p_{bf}(v_1,v_2)$ \\ 
         \hline
  0.1 & 0.2 & 0.188748693806651259$\times 10^{-2}$ & 3.3 & 3.4 & 0.707471726450662352$\times 10^{-2}$ & 11.4 & 11.5 & 0.132984011459287600$\times 10^{-4}$ \\
  0.2 & 0.3 & 0.599493924449560325$\times 10^{-2}$ & 3.4 & 3.5 & 0.630325324464037966$\times 10^{-2}$ & 11.8 & 11.9 & 0.108963869731728562$\times 10^{-4}$ \\
  0.3 & 0.4 & 0.119838730486052627$\times 10^{-1}$ & 3.5 & 3.6 & 0.562029109359469465$\times 10^{-2}$ & 12.2 & 12.3 & 0.898231826177633474$\times 10^{-5}$ \\
  0.4 & 0.5 & 0.191713037476076467$\times 10^{-1}$ & 3.6 & 3.7 & 0.501570335101668238$\times 10^{-2}$ & 12.6 & 12.7 & 0.744673944950352989$\times 10^{-5}$ \\
  0.5 & 0.6 & 0.268148704175592575$\times 10^{-1}$ & 3.7 & 3.8 & 0.448040726228595298$\times 10^{-2}$ & 13.0 & 13.1 & 0.620692664989309923$\times 10^{-5}$ \\
  0.6 & 0.7 & 0.342169825838334266$\times 10^{-1}$ & 3.8 & 3.9 & 0.400629926793153534$\times 10^{-2}$ & 13.4 & 13.5 & 0.519985882180620792$\times 10^{-5}$ \\
  0.7 & 0.8 & 0.408046817645270038$\times 10^{-1}$ & 3.9 & 4.0 & 0.358617903052689251$\times 10^{-2}$ & 13.8 & 13.9 & 0.437716091236265067$\times 10^{-5}$ \\
  0.8 & 0.9 & 0.461752111181489148$\times 10^{-1}$ & 4.0 & 4.1 & 0.321366933187318903$\times 10^{-2}$ & 14.3 & 14.4 & 0.355190209386678934$\times 10^{-5}$ \\
  0.9 & 1.0 & 0.501069962296554854$\times 10^{-1}$ & 4.1 & 4.2 & 0.288313611172121066$\times 10^{-2}$ & 14.7 & 14.8 & 0.301972189121095148$\times 10^{-5}$ \\
  1.0 & 1.1 & 0.525430066270895517$\times 10^{-1}$ & 4.4 & 4.5 & 0.209661698161469378$\times 10^{-2}$ & 15.0 & 15.1 & 0.268074181892024922$\times 10^{-5}$ \\
  1.1 & 1.2 & 0.535569109220308864$\times 10^{-1}$ & 4.5 & 4.6 & 0.188991759227905040$\times 10^{-2}$ & 15.3 & 15.4 & 0.238503756961240852$\times 10^{-5}$ \\
                     \hline                        
  1.2 & 1.3 & 0.533124182096042009$\times 10^{-1}$ & 4.6 & 4.7 & 0.170565123472654998$\times 10^{-2}$ & 15.5 & 15.6 & 0.220885480460924363$\times 10^{-5}$ \\
  1.3 & 1.4 & 0.520239546086940725$\times 10^{-1}$ & 4.8 & 4.9 & 0.139429327402048644$\times 10^{-2}$ & 16.0 & 16.1 & 0.183048927501968280$\times 10^{-5}$ \\
  1.4 & 1.5 & 0.499237380879169903$\times 10^{-1}$ & 5.0 & 5.1 & 0.126289351295904078$\times 10^{-2}$ & 17.0 & 17.1 & 0.127721410753661143$\times 10^{-5}$ \\
  1.5 & 1.6 & 0.472374282113263953$\times 10^{-1}$ & 5.2 & 5.3 & 0.945111931756639213$\times 10^{-3}$ & 18.0 & 18.1 & 0.908492839681844620$\times 10^{-6}$ \\
  1.6 & 1.7 & 0.441683859834397057$\times 10^{-1}$ & 5.4 & 5.5 & 0.783566193490159289$\times 10^{-3}$ & 19.0 & 19.1 & 0.657462602010086667$\times 10^{-6}$ \\
  1.7 & 1.8 & 0.408893070688341261$\times 10^{-1}$ & 5.6 & 5.7 & 0.652564255926359053$\times 10^{-3}$ & 20.0 & 20.1 & 0.483245270914358678$\times 10^{-6}$ \\
  1.8 & 1.9 & 0.375394578360157054$\times 10^{-1}$ & 5.8 & 5.9 & 0.545847733296032050$\times 10^{-3}$ & 21.0 & 21.1 & 0.360220839892367412$\times 10^{-6}$ \\
  1.9 & 2.0 & 0.342257171222356938$\times 10^{-1}$ & 6.0 & 6.1 & 0.458524742961601043$\times 10^{-3}$ & 22.0 & 22.1 & 0.271967525801591057$\times 10^{-6}$ \\
  2.0 & 2.1 & 0.310258812602125892$\times 10^{-1}$ & 6.7 & 6.8 & 0.257061914585502687$\times 10^{-3}$ & 23.0 & 23.1 & 0.207742414782007755$\times 10^{-6}$ \\
  2.1 & 2.2 & 0.279930533347722332$\times 10^{-1}$ & 7.0 & 7.1 & 0.203416661022484588$\times 10^{-3}$ & 24.0 & 24.1 & 0.160385605407817376$\times 10^{-6}$ \\
  2.2 & 2.3 & 0.251603010727732525$\times 10^{-1}$ & 7.3 & 7.4 & 0.162215738465301014$\times 10^{-3}$ & 25.0 & 25.1 & 0.125043165180335074$\times 10^{-6}$ \\
        \hline
  2.3 & 2.4 & 0.225450758213940055$\times 10^{-1}$ & 7.6  &  7.7 & 0.130309358504550521$\times 10^{-3}$ & 27.0 & 27.1 & 0.780383645690348706$\times 10^{-7}$ \\
  2.4 & 2.5 & 0.201531190401666584$\times 10^{-1}$ & 8.0  &  8.1 & 0.983551981869120013$\times 10^{-4}$ & 30.0 & 30.1 & 0.407187063104246108$\times 10^{-7}$ \\
  2.5 & 2.6 & 0.179817451095081113$\times 10^{-1}$ & 8.4  &  8.5 & 0.750897540252961992$\times 10^{-4}$ & 35.0 & 35.1 & 0.155222874488130268$\times 10^{-7}$ \\
  2.6 & 2.7 & 0.160224922926440151$\times 10^{-1}$ & 8.8  &  8.9 & 0.579397050271662135$\times 10^{-4}$ & 40.0 & 40.1 & 0.662851903623728573$\times 10^{-8}$ \\
  2.7 & 2.8 & 0.142631922581957772$\times 10^{-1}$ & 9.2  &  9.3 & 0.451506651248345601$\times 10^{-4}$ & 45.0 & 45.1 & 0.308047666225349931$\times 10^{-8}$ \\
  2.8 & 2.9 & 0.126895364261496964$\times 10^{-1}$ & 9.6  &  9.7 & 0.355100957492926310$\times 10^{-4}$ & 50.0 & 50.1 & 0.152690469597701984$\times 10^{-8}$ \\
  2.9 & 3.0 & 0.112862256268808421$\times 10^{-1}$ & 10.0 & 10.1 & 0.281690513621917630$\times 10^{-4}$ & 55.0 & 55.1 & 0.795413539717975718$\times 10^{-9}$ \\
  3.0 & 3.1 & 0.100377863252957321$\times 10^{-1}$ & 10.4 & 10.5 & 0.225258543290844444$\times 10^{-4}$ & 60.0 & 60.1 & 0.430580269208474099$\times 10^{-9}$ \\
  3.1 & 3.2 & 0.892912755022094856$\times 10^{-2}$ & 10.8 & 10.9 & 0.181491361268184092$\times 10^{-4}$ & 70.0 & 70.1 & 0.136754671307452372$\times 10^{-9}$ \\
  3.2 & 3.3 & 0.794590125773871769$\times 10^{-2}$ & 11.0 & 11.1 & 0.163344102596100687$\times 10^{-4}$ & 75.0 & 75.1 & 0.791005569608716576$\times 10^{-10}$ \\
        \hline \hline
  \end{tabular}}
  \end{center}
  \end{table}
\begin{table}[tbp]
   \caption{Numerical values of the inverse normalization constants, i.e. the $N^{-1}_{\gamma} = \Bigl({\cal C}^{\prime}_{\gamma}\Bigr)^{-1}$ value in the spectral 
            function for the primary $\beta^{-}-$electrons, Eq.(\ref{eq555a}), as the function of the thermal effect $\Delta E^{\prime}$ (in $MeV$) of the nuclear 
            $\beta^{-}$-decay in light atoms.} 
     \begin{center}
     \scalebox{0.80}{%
     \begin{tabular}{| c | c | c | c | c | c |}
      \hline\hline
 $\Delta E^{\prime}$ & $N^{-1}_{\gamma}$ & $\Delta E^{\prime}$ & $N^{-1}_{\gamma}$  & $\Delta E^{\prime}$ & $N^{-1}_{\gamma}$ \\ 
         \hline
 0.6 & 0.510138920728380682$\times 10^{-3}$ & 2.6 & 0.943534198627184360$\times 10^{2}$ & 4.6 & 0.185482676760179721$\times 10^{4}$ \\
 0.7 & 0.768274953852744202$\times 10^{-2}$ & 2.7 & 0.115479629241228239$\times 10^{3}$ & 4.7 & 0.207060009101156116$\times 10^{4}$ \\
 0.8 & 0.365273098535936243$\times 10^{-1}$ & 2.8 & 0.140174066944844508$\times 10^{3}$ & 4.8 & 0.230589234934879363$\times 10^{4}$ \\
 0.9 & 0.110787349453086509                 & 2.9 & 0.168863421848105529$\times 10^{3}$ & 4.9 & 0.256199131172405105$\times 10^{4}$ \\
 1.0 & 0.263733079531422292                 & 3.0 & 0.202006532798188682$\times 10^{3}$ & 5.0 & 0.284023985445725021$\times 10^{4}$ \\
 1.2 & 0.992531073989599566                 & 3.2 & 0.283657013608144685$\times 10^{3}$ & 5.2 & 0.346883961029256005$\times 10^{4}$ \\
 1.4 & 0.271944662442253340$\times 10^{1}$  & 3.4 & 0.389482828970674785$\times 10^{3}$ & 5.4 & 0.420358039547922170$\times 10^{4}$ \\
 1.5 & 0.417197363819929210$\times 10^{1}$  & 3.5 & 0.452982885446960540$\times 10^{3}$ & 5.5 & 0.461472909694107252$\times 10^{4}$ \\
 1.6 & 0.616262631954181568$\times 10^{1}$  & 3.6 & 0.524429148588098765$\times 10^{3}$ & 5.6 & 0.505730617602198678$\times 10^{4}$ \\
 1.7 & 0.882174671793232731$\times 10^{1}$  & 3.7 & 0.604537642065602785$\times 10^{3}$ & 5.7 & 0.553307240355429714$\times 10^{4}$ \\
 1.8 & 0.122980724837757482$\times 10^{2}$  & 3.8 & 0.694065721429933329$\times 10^{3}$ & 5.8 & 0.604385284163182761$\times 10^{4}$ \\
 1.9 & 0.167598801452295380$\times 10^{2}$  & 3.9 & 0.793813221987212581$\times 10^{3}$ & 5.9 & 0.659153799162885377$\times 10^{4}$ \\
 2.0 & 0.223961294672658384$\times 10^{2}$  & 4.0 & 0.904623606693807949$\times 10^{3}$ & 6.0 & 0.717808494222105881$\times 10^{4}$ \\
 2.2 & 0.380580798087427918$\times 10^{2}$  & 4.2 & 0.116303190610637340$\times 10^{4}$ & 6.2 & 0.847593242453915958$\times 10^{4}$ \\
 2.4 & 0.612527955319830414$\times 10^{2}$  & 4.4 & 0.147695449723875506$\times 10^{4}$ & 6.4 & 0.995442736892640637$\times 10^{4}$ \\
 2.5 & 0.763997524066278397$\times 10^{2}$  & 4.5 & 0.165733856922900762$\times 10^{4}$ & 6.5 & 0.107670505698641369$\times 10^{5}$ \\
        \hline \hline
  \end{tabular}}
  \end{center}
  \end{table}

\begin{thebibliography}{01}

\bibitem{Blat} J.M. Blatt and V.F. Weisskopf, \textit{Theoretical Nuclear Physics}, (Springer-Verlag Inc., New York (1979)). 

\bibitem{Fro05} A.M. Frolov and J.D. Talman, Phys. Rev. A {\bf 72}, 022511 (2005) [ ibid, {\bf 57}, 2436 (1998) (about $\beta^{-}$ decay of the T$^{-}$ ion)].

\bibitem{Mig1940} A.B. Migdal, J. Phys. (Moscow) {\bf 4}, 441, (1941).

\bibitem{Our1} A.M. Frolov and M.B. Ruiz, Phys. Rev. A {\bf 82}, 042511 (2010) [see also: Adv. Quant. Chem. {\bf 67}, 267 (2013)].

\bibitem{Fro98} A.M. Frolov, Phys. Rev. A {\bf 57}, 2436 (1998) [ibid, {\bf 79}, 032703 (2009)].

\bibitem{PRC1} F. Simkovic, R. Dvornicky and A. Faessler, Phys. Rev. C {\bf 77}, 055502 (2008). 

\bibitem{PRC2} N. Doss, J. Tennyson, A. Saenz and S. Jonsell, Phys. Rev. C {\bf 73}, 025502 (2006). 

\bibitem{LLQ} L.D. Landau and E.M. Lifshitz, {\it Quantum Mechanics: non-relativistic theory}, (3rd. ed. Pergamon Press, New York (1976)), Chpt. VI. 

\bibitem{MigK} A.B. Migdal and V. Krainov, {\it Approximation Methods in Quantum Mechanics}, (W.A. Benjamin, New York (1969)).

\bibitem{Maxim} H.A. Bethe and L.C. Maximon, Phys. Rev. {\bf 93}, 768 (1954).

\bibitem{GR} I.S. Gradstein and I.M. Ryzhik, \textit{Tables of Integrals, Series and Products}, (6th revised ed., Academic Press, New York (2000)).

\bibitem{AS} \textit{Handbook of Mathematical Functions} (M. Abramowitz and I.A. Stegun (Eds.), Dover, New York, 1972).

\bibitem{Fro2015} A.M. Frolov, Letters to JETP {\bf 103}, 173 (2016) [JETP Letters {\bf 103}, 173 (2016)]. 

\bibitem{Jack} J.D. Jackson, \textit{Classical Electrodynamics}, (2nd ed., J. Wiley and Sons Inc., New York (1975)).

\bibitem{March} N.H. March, W.H. Young and S. Sampanthar, \textit{The Many-Body Problem in Quantum Mechanics}, (Dover Publ. Inc., New York (1995)).

\bibitem{Fro06} A.M. Frolov, J. Phys. A {\bf 39}, 15421 (2006).

\bibitem{Dirac} P.A.M. Dirac, Proc. Camb. Phil. Soc. {\bf 34}, 204 (1930). 

\bibitem{MCQ} R. McWeeny and B.T. Sutcliffe, \textit{Methods of Molecular Quantum Mechanics}, (Academic Press, New York (1969)).

\bibitem{David} E.R. Davidson, Adv. Quant. Chem. {\bf 6}, 235 (1972).

\bibitem{Recp} W.H. Press, S.A. Teulkolsky, W.T. Vetterhg and B.P. Flannery, \textit{Numerical Recipes in FORTRAN} (2nd edn, Cambridge University Press, Cambbridge UK, 1996). 

\bibitem{Num} R.L. Burden, J. Douglas Faires and A.C. Reynolds, {\it Numerical Analysis}, (2nd ed., PWS Publishers, Boston, MA, 1981).

\bibitem{Fro2016} A.M. Frolov, Europhysics Letters (EPL) {\bf 133} 12001 (2016). 
       
\bibitem{Fermi} E. Fermi, Zeits. f\"{u}r Physik {\bf 88}, 172 (1934).

\bibitem{Long} C. Longmire and H. Brown, Phys. Rev. {\bf 75}, 264 (1949).

\bibitem{Hall} H. Hall, Phys. Rev. {\bf 79}, 745 (1950).

\bibitem{Mei} J.Y. Mei, Phys. Rev. {\bf 81}, 287 (1951).

\bibitem{Konop} E.J. Konopinski, Rev. Mod. Phys. {\bf 15}, 209 (1943).

\bibitem{Cook} L.S. Cook and L.M. Langer, Phys. Rev. {\bf 73}, 601 (1948).

\bibitem{Neary} G. J. Neary, Roy. Phys. Soc. (London), {\bf A175}, 71 (1940).

\bibitem{Bethe} H.A. Bethe and R.F. Bacher, Rev. Mod. Phys. {\bf 8}, 82 (1935).

\end{thebibliography}
\end{document}